\documentclass[prl,reprint,a4paper,floatfix,superscriptaddress]{revtex4-1}

\usepackage{graphicx}
\usepackage{amsmath}
\usepackage{hyperref}
\usepackage{url}
\usepackage{epstopdf}

\newcommand{\musr}{$ \mu $SR }
\newcommand{\tnone}{ T^{\mathrm{N1}} }
\newcommand{\tntwo}{ T^{\mathrm{N2}} }
\newcommand{\tnthree}{ T^{\mathrm{N3}} }
\newcommand{\bmag}{\mathbf{B}_{\mathrm{mag}}}
\newcommand{\bext}{\mathbf{B}_{\mathrm{ext}}}
\newcommand{\bmu}{\mathbf{B}_{\mu}}
\newcommand{\vmu}{\nu_{\mu}}
\newcommand{\pmu}{\mathbf{P}_{\mu}}
\newcommand{\mub}{\mu_{\mathrm{B}}}

\begin{document}

\title{Muon-spin-rotation study of the magnetic structure in the tetragonal antiferromagnetic state of weakly underdoped Ba$ _{1-x} $K$ _{x} $Fe$ _{2} $As$ _{2} $}

\author{B.P.P. Mallett}
\affiliation{University of Fribourg, Department of Physics and Fribourg Center for Nanomaterials, Chemin du Mus\'{e}e 3, CH-1700 Fribourg, Switzerland}

\author{Yu. G. Pashkevich}
\affiliation{O. O. Galkin Donetsk Institute for Physics and Engineering NAS of Ukraine, 03680 Kyiv, Ukraine}

\author{A. Gusev}
\affiliation{O. O. Galkin Donetsk Institute for Physics and Engineering NAS of Ukraine, 03680 Kyiv, Ukraine}

\author{Th. Wolf}
\affiliation{Institute of Solid State Physics, Karlsruhe Institute of Technology, Postfach 3640, Karlsruhe 76021, Germany}

\author{C. Bernhard}
\affiliation{University of Fribourg, Department of Physics and Fribourg Center for Nanomaterials, Chemin du Mus\'{e}e 3, CH-1700 Fribourg, Switzerland}

\date{\today}


\begin{abstract}

With muon spin rotation ($ \mu $SR) we studied the transition between the orthorhombic antiferromagnetic (o-AF) and the tetragonal antiferromagnetic (t-AF) states of a weakly underdoped Ba$ _{1-x} $K$ _{x} $Fe$ _{2} $As$ _{2} $ single crystal. We observed some characteristic changes of the magnitude and the orientation of the magnetic field at the muon site which, due to the fairly high point symmetry of the latter, allow us to identify the magnetic structure of the t-AF state. It is the so-called, inhomogeneous double-$\mathbf{Q}$ magnetic structure with $ c $-axis oriented moments which has a vanishing  magnetic moment on half of the Fe sites.

\end{abstract}

\maketitle

The phase diagram of the iron arsenide superconductors is characterized by the close proximity of the antiferromagnetic (AF) and superconducting (SC) orders \cite{paglione2010,stewart2011,lumsden2010,chubukov2012,mazin2008}. In the underdoped regime they coexist and compete for the same low-energy electronic states \cite{pratt2009,marsik2010}. Moreover, the SC transition temperature, $ T_{\mathrm{c}} $, reaches its maximum near the point where the AF order vanishes. It is therefore commonly assumed that the AF fluctuations are involved in the SC pairing mechanism \cite{chubukov2012,mazin2008}. Accordingly, it is important to obtain a full understanding of the AF ground state and also of the nearly degenerate states which can strongly impact the magnetic fluctuations.

For the 122-type BaFe$ _{2} $As$ _{2} $ system, the primary AF order is a so-called single-$\mathbf{Q}$ state with in-plane antiparallel spins along the $ (0, \pi) $ direction and parallel ones along $ (\pi, 0) $ \cite{paglione2010,stewart2011,lumsden2010,delacruz2008}. The transition to this so-called ``stripe-like'' AF state is coupled with a structural transition from a tetragonal paramagnetic to an orthorhombic AF (o-AF) state. The interactions underlying this magneto-structural transition are the subject of an ongoing discussion. The explanations range from the so-called itinerant models for which the Fermi-surface nesting governs the magnetic interactions \cite{chubukov2012,mazin2008,fernandes2014} to orbital models which assume localized spins whose exchange interactions are determined by the orbital occupation \cite{lee2009, kruger2009}. An important role of the spin-lattice coupling \cite{yildirim2009} and a near degeneracy of different spin states of the Fe ions have also been discussed \cite{chaloupka2013, gretarsson2013}.

In view of this ongoing discussion, the recent observation of a phase transition from the o-AF state to a new type of AF state with tetragonal lattice symmetry (t-AF) has attracted considerable attention.  This t-AF state was first observed at ambient pressure in Ba$ _{1-x} $Na$ _{x} $Fe$ _{2} $As$ _{2} $ (BNFA) \cite{avci2014, wasser2015} and more recently in Ba$ _{1-x} $K$ _{x} $Fe$ _{2} $As$ _{2} $ (BKFA) \cite{bohmer2014, allred2015xray} in a narrow doping range close to optimal doping. A state with very similar phenomenology was observed earlier in BKFA at elevated pressures \cite{hassinger2012}. In BKFA, even a reentrance of the o-AF state at lower temperature has been observed which seems to be induced (or at least supported) by superconductivity \cite{bohmer2014,mallett2015bkfaIR}.

Motivated by these observations, a range of so-called double-$\mathbf{Q}$ magnetic structures have been proposed for the t-AF state \cite{kang2015,*wang2015, *fernandes2015, khalyavin2014, gastiasoro2015}. These correspond to different superpositions of the single-$\mathbf{Q}$ states with wave vectors $ (0, \pi) $ and $ (\pi, 0) $. A perpendicular magnetisation direction of the single-$\mathbf{Q}$ components yields the so-called orthomagnetic states whilst a parallel one results in a non-uniform state with vanishing Fe moments on half of the lattice sites \cite{kang2015,*wang2015, *fernandes2015, khalyavin2014, gastiasoro2015, giovannetti2011}. 

From neutron diffraction it is known that the o-AF to t-AF transition is accompanied by a spin reorientation from the in-plane to the out-of-plane direction \cite{wasser2015, allred2015xray}. However, it is not possible to distinguish between the different double-$\mathbf{Q}$ structures, and even a twinned state of a single-$\mathbf{Q}$ structure with c-axis oriented moments cannot be excluded \cite{wasser2015}. 

In this letter we present a muon spin rotation ($ \mu $SR) study of the o-AF to t-AF transition and its reversal in the SC state of underdoped Ba$ _{1-x} $K$ _{x} $Fe$ _{2} $As$ _{2} $. In particular, we show that the changes of the magnitude and the orientation of the magnetic field at the muon site are characteristic of the non-uniform double-$\mathbf{Q}$ magnetic structure for which half of the Fe sites have zero (or at least a very small) magnetic moment. 

\begin{figure}
		\includegraphics[width=1.0\columnwidth]{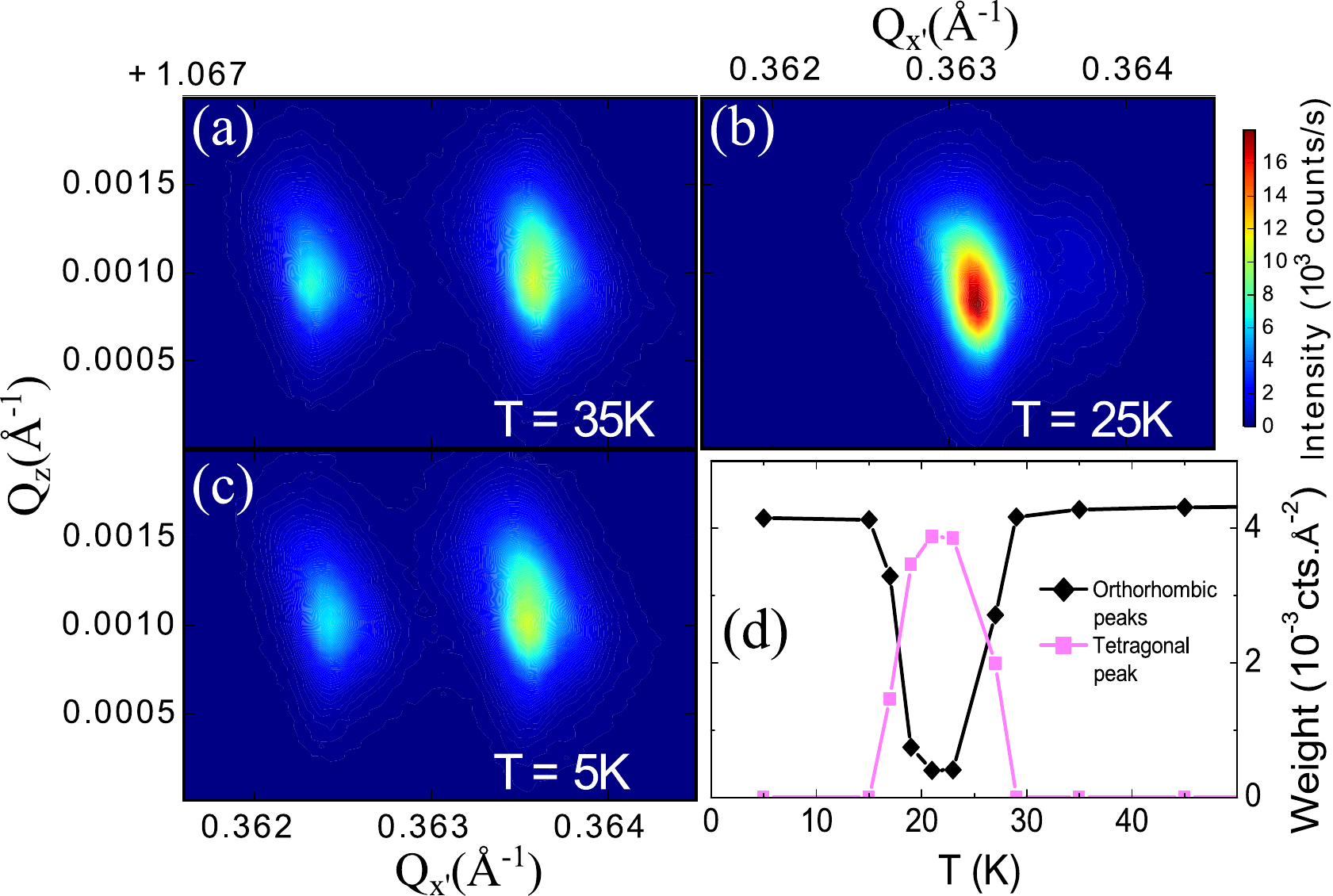}
	\caption{\label{fig1} Diffracted x-ray intensity in a region of reciprocal space around the $ (1,1,14) $ Bragg-peak (a) in the o-AF state between $ \tnone $ and $ \tntwo $, (b) in the t-AF state and (c) in the o-AF state below $ \tnthree $. (d) The temperature dependence of the combined weight of the split orthorhombic peaks and the weight of the tetragonal peak.}
\end{figure} 
 
A crystal with an $ ab $ surface of about $ 2 \times 4  $~mm$ ^{2} $ and a thickness close to 100~$ \mu $m was grown in an alumina crucible using FeAs flux as described in Ref.~\cite{karkin2014}. For a different crystal from the same batch the K-content has been determined with x-ray diffraction refinement to be $ x=0.247(2) $. Other crystals from this batch have been investigated with thermal expansion and specific heat \cite{bohmer2014} as well as with neutron diffraction which confirm the $ c $-orientation of the spins in the t-AF state \cite{suprivate}. The very same crystal has also been studied with resistivity, magnetisation and infrared measurements which revealed the onset of a weak SC state at $ T_{\mathrm{c}}^{\mathrm{ons}} \approx 28 $~K and a second transition to a stronger SC state at 18~K \cite{mallett2015bkfaIR}. The structural transitions have been investigated with a 4-circle x-ray diffractometer (RIGAKU SmartLab) with a 9 kW rotating anode Cu K$ \alpha $ source and a He-flow cryostat. 
Figure~\ref{fig1}(a-c) shows the diffracted x-ray intensity in a region of reciprocal space around the $ (1,1,14) $ Bragg-peak (in the $ I4/mmm $ representation) for selected temperatures. The Bragg-peak exhibits a sizeable splitting in the $ (1,1,0) $ direction (denoted by $ Q_{x'} $) in the o-AF state below $ \tnone \approx 72$~K, that starts to disappear below $ \tntwo \approx 32 $~K, and reappears below $ \tnthree \approx 18 $~K. A quantitative analysis of the Bragg-peaks, as shown in Refs. \cite{mallett2015bkfaIR, mallett2015bkfamusrSOM} and Fig.~\ref{fig1}(d), confirms that in the t-AF state only a minor fraction,  about 10\%, remains in the o-AF phase.

The \musr measurements were performed at the GPS instrument of the $ \mu $M3 beamline at the Paul Scherrer Institute (PSI) in Villigen, Switzerland. Fully spin-polarized, positive muons with an energy of $ 4.2 $~MeV were implanted in the crystal (along the $ c $-axis of the crystal) where they rapidly thermalize and stop at interstitial lattice sites distributed over a depth of about 100~$ \mu $m. The muon spins precess in the magnetic field at the muon site, $ \bmu $, with a precession frequency $ \vmu = \gamma_{\mu}\bmu/2\pi $, where $ \gamma_{\mu}=2\pi\times135.5 $~MHz/T is the gyromagnetic ratio. The time evolution of the polarization of the muon spin ensemble, $ P(t) $, is detected via the asymmetry of the emission rate of the decay positrons as described in Refs. \cite{schenck1985,lee1999}. The zero-field (ZF) and transverse-field (TF) measurements were performed with the so-called up-down positron counters in spin rotation mode for which the muon spin polarisation, $ \pmu $, is at about 54$ ^{\circ} $ with respect to the muon beam (pointing toward the upward counter), see Fig.~S7(b) of Ref.~\cite{mallett2015bkfamusrSOM}.

 \begin{figure}

 		\includegraphics[width=1.0\columnwidth]{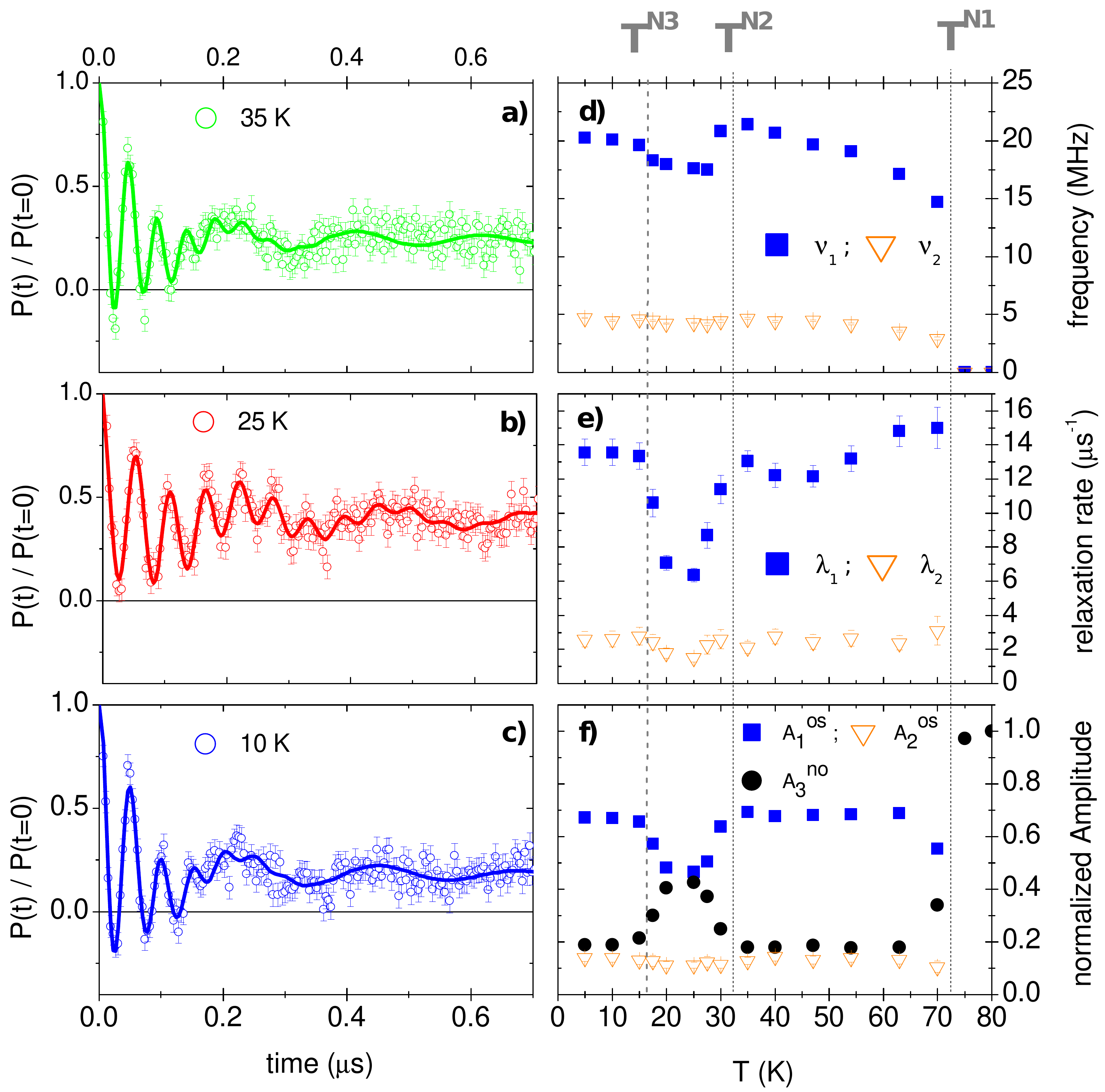}
 	\caption{\label{fig2} ZF-\musr curves in (a) the o-AF, (b) the t-AF and (c) the reentrant o-AF states. Temperature dependence of (d) the muon precession frequencies, (e) the relaxation rates and (f) the amplitudes of the oscillating and non-oscillating components.}
 	
 \end{figure} 

Figure~\ref{fig2} shows representative ZF-\musr spectra (symbols) in (a) the o-AF state at $ T<\tnone $, (b) the t-AF state at $ \tnthree< T <\tntwo $, and (c) after the reentrance into the o-AF state at $ T<\tnthree $. The solid lines show the fits with the function
\[
 P(t) = P(0)\sum_{i=1}^{2}{A_{i}^{\mathrm{os}} \cos{(\gamma_{mu}B_{\mu,i}t+\phi_{i})e^{-\lambda_{i}t} }
+ A_{3}^{\mathrm{no}}e^{-\lambda_{3}t} } 
\] 
\noindent where the parameters $ P $, $ A_{i} $ ,$ B_{\mu,i} $, $ \phi_{i} $, and $ \lambda_{i} $ describe the polarization of the muon spin ensemble, the relative amplitudes of the signals, the magnitude of the local magnetic field at the muon sites, the initial phase of the muon spin, and the relaxation rates, respectively. In agreement with previous studies of undoped (Sr,Ba)Fe$ _{2} $As$ _{2} $ \cite{jesche2008,aczel2008,bernhard2012} and underdoped Ba$ _{1-x} $K$ _{x} $Fe$ _{2} $As$ _{2} $ \cite{wiesenmayer2011} we distinguish the precessing signals from two muon sites with inequivalent fields, $ \bmu $, via two distinct signals with different precession frequencies, $ \nu_{\mu,1,2} $, and  amplitude fractions, $ A_{1,2}^{\mathrm{os}} $. The non-precessing signal described by the third term arises from the non-orthogonal orientation of $ \pmu $ and $ \bmu $ and also from a small background due to muons that stop outside the sample.

\begin{figure*}

		\includegraphics[width=0.85\textwidth]{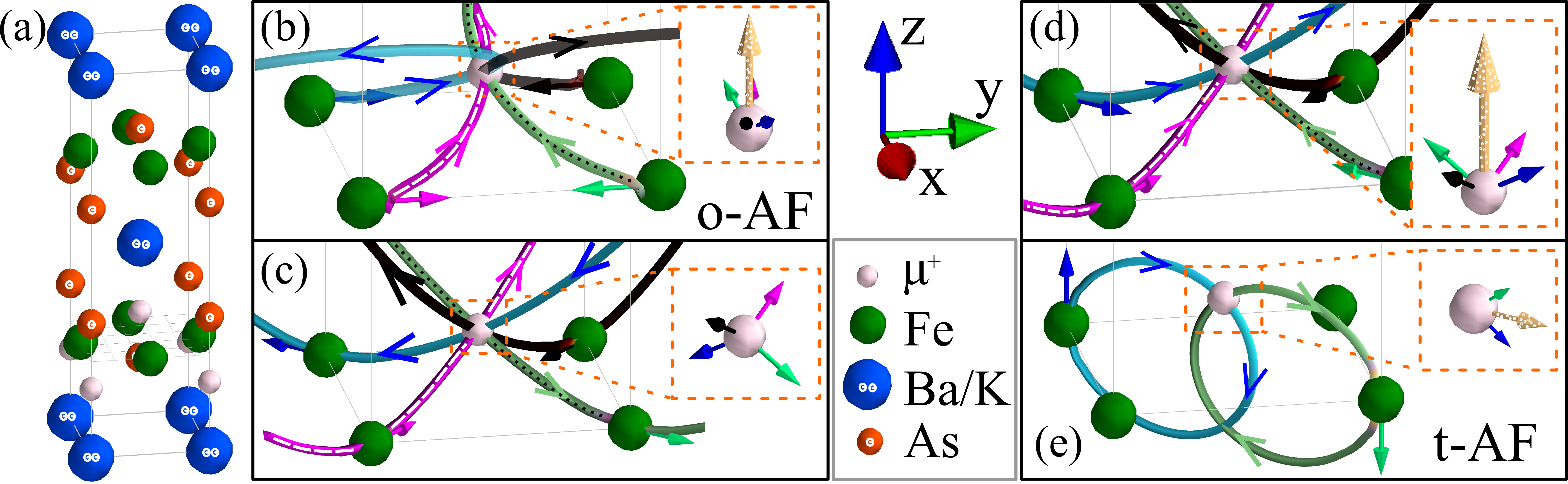}
	\caption{\label{fig3} (a) The structure of Ba$ _{1-x} $K$ _{x} $Fe$ _{2} $As$ _{2} $ showing the high symmetry of the main muon site. (b-e) The contributions from the Fe magnetic moments to the dipole field at the muon site, $ \bmu $, for the different orders. The insets show how the dipole fields  add or cancel to produce $ \bmu $ (dotted yellow arrow). (b) The single-$\mathbf{Q}$ structure found in the o-AF state. (c) Orthomagnetic, double-$\mathbf{Q}$ structure which yields $ \bmu=0 $. Orthomagnetic double-$\mathbf{Q}$ structure where $ \bmu $ is larger than that in the o-AF state. (e) The non-uniform double-$\mathbf{Q}$ structure that is consistent with the experiment.}
	
\end{figure*} 

To calculate the muon site we have used a modified Thomas Fermi approach \cite{reznik1995} with the structural data as specified in Ref.~\cite{mallett2015bkfamusrSOM}. This approach has already been used to successfully predict the muon sites in related magnetic materials such as RFeAsO \cite{maeter2009}, RFeO$ _{3} $ \cite{holzschuh1983} and RBaCo$ _{2} $O$ _{5.5} $ \cite{luetkens2008}. As shown in Fig.~\ref{fig3}(a), the majority muon site due to the global minimum in the potential energy is located at the coordinate $ (0, 0, 0.191) $ in the $ I4/mmm $ setting, i.e. on the line that connects the Ba and As ions along the $ c $-direction. The corresponding site in the vicinity of the K ions is $ (0, 0, 0.17) $ \cite{mallett2015bkfamusrSOM}. Dipolar field calculations show that these muon sites have nearly identical local fields, $ \bmu $, and thus both contribute to the signal with $ \vmu(T\!\rightarrow\! 0)\approx 21 $~MHz. A similar muon site was previously found in RFeAsO \cite{maeter2009}. The oscillatory signal with the smaller amplitude, $ A_{2}^{\mathrm{os}} $, and $ \vmu(T\!\rightarrow\! 0)\approx 4 $~MHz originates from a secondary muon site due to a local potential minimum. It is located at $ (0.4, 0.5, 0) $, slightly away from the line connecting the As ions along the $ c $-direction. It also has a high local symmetry with the same direction and qualitative change of the local magnetic field. In the following discussion we focus on the majority muon site and the change of its local field during the o-AF to t-AF transition. The high point symmetry of this muon site (the same as for the As ions) helps us to distinguish between the different double-$\mathbf{Q}$ magnetic structures that have been proposed for the t-AF state. Note that the partially itinerant nature of the magnetic moments does not invalidate the symmetry based arguments used.

Figures~\ref{fig2}(d)-(f) display the $ T $-dependence of the fit parameters of the ZF-\musr spectra which exhibit characteristic changes at the transition from the o-AF to the t-AF state. These involve a moderate reduction of $ \nu_{\mu,1}=\frac{\gamma_{\mu}}{2\pi} B_{\mu,1}$, a sizeable decrease (increase) of $ A_1 $ ($ A_3 $), and a surprisingly large reduction of $ \lambda_{1} $. The latter can be understood in terms of the disorder from the magnetic domain walls of the single-$\mathbf{Q}$ order in the o-AF state that are absent for the double-$\mathbf{Q}$ structure in the t-AF state. The former effects indicate that the magnitude of $ \bmu $ decreases only moderately whereas its direction changes substantially at the o-AF to t-AF transition \cite{mallett2015bkfamusrSOM}.

\begin{figure}

		\includegraphics[width=0.65\columnwidth]{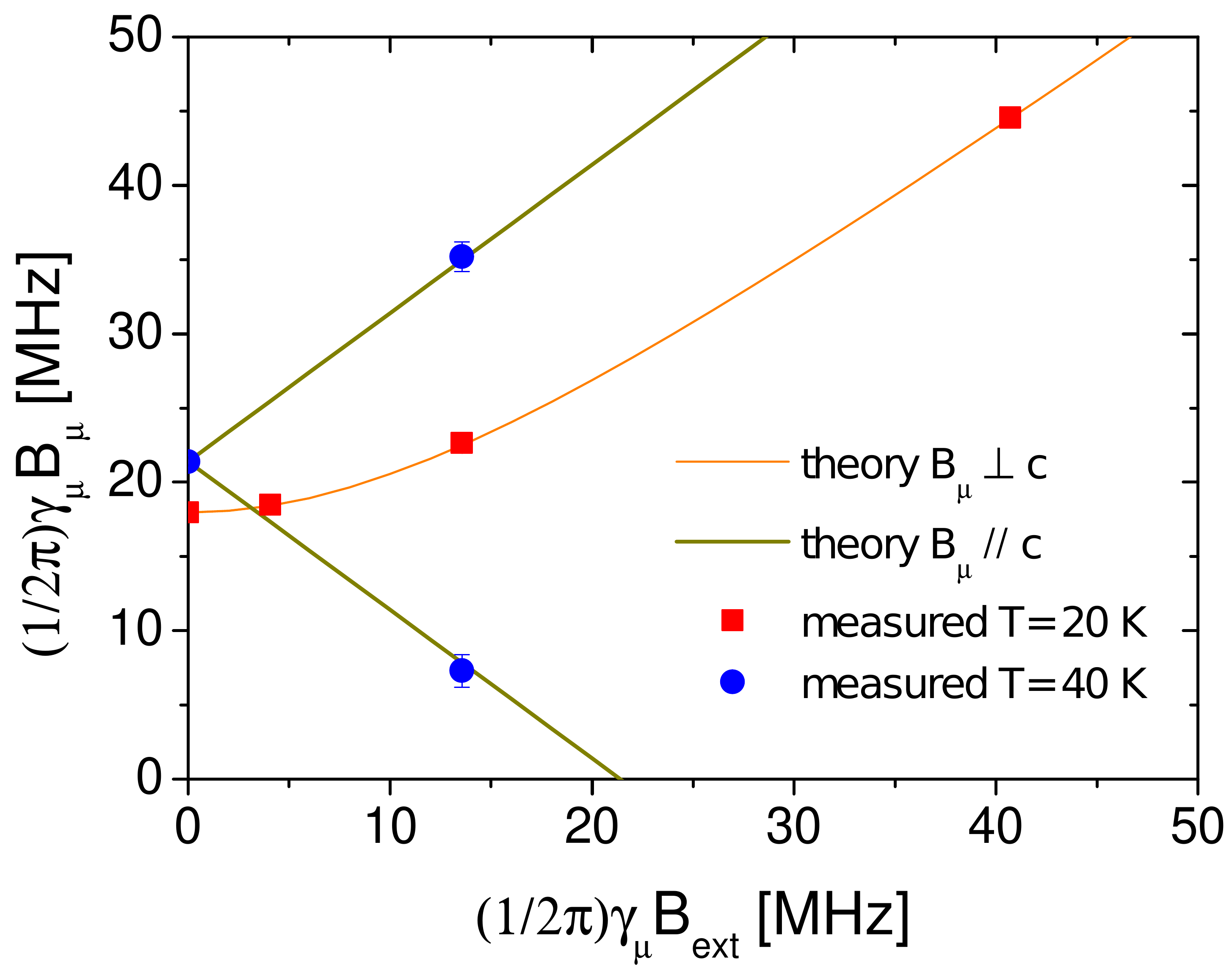}
	\caption{\label{fig4} TF-\musr data showing the measured local magnetic field versus the external field that is applied in the $ c $-direction.}
	
\end{figure}

For the single-$\mathbf{Q}$ order in the o-AF state our calculations predict that, whilst the Fe moments are orientated in the plane, $ \bmu $ is parallel to the $ c $-axis, as shown in Fig.~\ref{fig3}(b) and Ref. \cite{mallett2015bkfamusrSOM}. This scenario is confirmed by the TF-\musr data  shown in Fig.~\ref{fig4} and also by the ZF-\musr data using the forward and backward counters that are discussed in Ref. \cite{mallett2015bkfamusrSOM}. Assuming a magnetic moment of 1~$ \mub $/Fe ion, the calculation for the single-$\mathbf{Q}$ structure in the o-AF state yields a precession frequency of about $ \vmu\approx37 $~MHz.  For undoped Ba-122, with a moment of about 1~$ \mub $/Fe as obtained from neutron scattering (see table 2 in Ref.~\cite{lumsden2010}), the \musr experiments yield a value of $ \vmu\approx29 $~MHz \cite{aczel2008,bernhard2012}. The latter is about 20\% lower than the value from the dipolar field calculations. In the following we assume that this 20\% difference between the calculated and the measured values of $ \vmu $ occurs also for our present sample. Accordingly, from the experimental value of $ \vmu\approx21 $~MHz in the o-AF state we obtain an estimate of the magnetic moment of about $ 0.7 $~$ \mub $/Fe.

Now we consider how $ \bmu $ and thus $ \vmu $ change for the different, predicted magnetic structures of the t-AF state that are shown in Figs.~\ref{fig3}(c-e) and detailed in Ref.~\cite{mallett2015bkfamusrSOM}. First of all, we can exclude the scenario discussed in Ref.~\cite{wasser2015}, that the magnetic order in the t-AF state has a twinned, single-$\mathbf{Q}$ structure similar as in the o-AF state but with the spins oriented along the $ c $-axis (as opposed to the in-plane orientation in the o-AF state). Our dipole field calculations in Ref.~\cite{mallett2015bkfamusrSOM} show that such a spin reorientation would merely affect the direction of $ \bmu $ but would not change its magnitude and thus $ \vmu $. This disagrees with our \musr data for which $ \vmu $ exhibits a clear decrease from 21 to 18~MHz at the o-AF to t-AF transition. 

Next, we address the non-collinear double-$\mathbf{Q}$ structures. As outlined in Ref.~\cite{mallett2015bkfamusrSOM}, these can be divided into two categories according to their symmetry. For the first category there is a complete cancellation of the dipolar field such that $ \vmu $ is zero, as shown in Fig.~\ref{fig3}(c) and in Ref.~\cite{mallett2015bkfamusrSOM}. For the second category, as shown in Fig.\ref{fig3}(d) and Ref.~\cite{mallett2015bkfamusrSOM}, the local field becomes very large and, in fact, significantly exceeds the one in the o-AF state. Assuming a constant moment of $ 0.7 $~$ \mub $/Fe ion, it would lead to an increase of $ \vmu $ from 21~MHz in the o-AF state to about 29~MHz in the t-AF state. For both categories the expected changes of $ \vmu $ at the o-AF to t-AF transition are thus entirely inconsistent with the observed moderate decrease of $ \vmu $. Furthermore, for the latter category the predicted orientation of $ \bmu $ along the $ c $-axis would also be in contradiction with the experimental result that $ \bmu $ is parallel to the $ ab $-plane, as shown below in Fig.~\ref{fig4}.

Finally, we show that our \musr data agree well with the prediction for the non-uniform double-$\mathbf{Q}$ structure with zero spin on half of the Fe sites and a sizeable, $ c $-axis oriented moment on the others (see Fig.~\ref{fig3}(e)). Assuming once more a moment of $ 0.7 $~$ \mub $/Fe ion as in the o-AF state, our calculations yield a reduction of $ \vmu $ from 21~MHz in the o-AF state to $ 14.8 $~MHz in the t-AF state. A quantitative agreement with the experimentally observed decrease of $ \vmu $ from 21 to 18~MHz is obtained if one assumes that the Fe moment (on the magnetic sublattice)  increases to $ 0.85 $~$ \mub $ in the t-AF state. Evidence for such an increase of the Fe magnetic moment in the t-AF state has indeed been obtained from infrared spectroscopy \cite{mallett2015bkfaIR}.

Our \musr data also agree well with the predicted change of the direction of $ \bmu $ from the $ c $-axis orientation in the o-AF state (as discussed above) to an in-plane orientation in the t-AF state, see Fig.~\ref{fig3}(e). This has been verified with a TF-\musr experiment for which the external field, $ \bext $ is applied along the $ c $-direction. The field at the muon site, $ \bmu $, is now a superposition of the fields due to the ordered moments, $ \bmag $, and of $ \bext $. For $ \bmag \perp \bext$ this yields the relationship $ \vmu = \frac{\gamma_{\mu}}{2\pi}  \sqrt{\bmag^{2}+\bext^{2}} $ which, as shown in Fig.~\ref{fig4}, the experimental data in the t-AF state follow very closely. This is clearly different from the behavior in the o-AF state, where $ \bmag $   is parallel or antiparallel to $ \bext $ ($ \bmag $   is pointing away from the nearest Fe plane), and the data are described by the relationship $ \vmu = \frac{\gamma_{\mu}}{2\pi} \left( \bmag \pm \bext \right) $ as was previously also found in Ref.~\cite{maeter2012}.

The identification of the magnetic order in the t-AF state in terms of the inhomogeneous double-$\mathbf{Q}$ structure raises the question whether the magnetic moment is strictly zero for half of the Fe sites or whether it is just smaller than on the other Fe sub-lattice. In Ref.~\cite{mallett2015bkfamusrSOM} we show that a corresponding magnetic structure with finite but inequivalent magnetic moments on both Fe sublattices could also account for the observed decrease of $ \vmu $. However, such a finite ordered moment would break the $ C_4 $ symmetry and thus should give rise to an orthorhombic lattice distortion similar to the one in the o-AF state. This is in contrast to the x-ray data in Fig.~\ref{fig1}(b) and Ref.~\cite{mallett2015bkfamusrSOM} which show no evidence of an additional broadening or splitting of the $ (1, 1, 14) $ Bragg peak in the t-AF state. A rough estimate shows that such a distortion must be at least two orders of magnitude smaller than the one in the o-AF state \cite{mallett2015bkfamusrSOM}. This conclusion is supported by a very recent M\"{o}ssbauer study which finds that about 50\% of the Fe sites do not host a measurable ordered magnetic moment \cite{allred2015}. While the authors of Ref.~\cite{allred2015} discuss their results mainly in terms of the itinerant magnetic model, we would like to mention the implications of a scenario whereby the Fe moments are partially localized. This scenario would suggest a variation of the spin state of the Fe ions from a low-spin state on the non-magnetic sublattice to an intermediate or high spin state on the magnetic one. The related change in the 3d orbital occupation may explain the spin reorientation. It should also induce some lattice distortions. An alternation of the Fe-As bond lengths is prohibited for the edge-sharing geometry of the Fe-As-tetrahedra but a displacement of the Fe ions away from their central position within the Fe-As tetrahedral is possible. The latter maintains the tetragonal $ C_4 $ symmetry but gives rise to a 4-fold increase of the unit cell size. Evidence for the resulting, additional IR-active modes in the t-AF state has indeed been observed with infrared spectroscopy \cite{mallett2015bkfaIR}. A variation of the spin state degree of freedom thus may need to be considered in a theoretical description of the magnetic properties of these iron arsenides \cite{chaloupka2013}, especially if one wants to account for the spin fluctuations which can be strongly influenced by the presence of nearly degenerate magnetic states with different spin states. 

In summary, we have studied with muon spin rotation how the local magnetic field at the muon site, $ \bmu $, changes at the transition from the o-AF to the t-AF state in a weakly underdoped BKFA single crystal. We have observed characteristic changes of the magnitude and the orientation of $ \bmu $ which enabled us to identify the magnetic order in terms of the inhomogeneous, double-$\mathbf{Q}$ magnetic structure for which the magnetic moment vanishes on every second Fe site. The unambiguous assignment of this magnetic structure has been possible thanks to the high point symmetry of the muon site.

\begin{acknowledgments}
This work was supported by the Schweizerische Nationalfonds (SNF) through grant No. 200020-153660. YuGP acknowledges support from NAS of Ukraine Grants No.78-02-14 
and  No. 53/15-N. We thank Christoph Meingast and Fred Hardy for stimulating discussions and Yixi Su for sharing with us his unpublished neutron diffraction data. Part of this work has been performed at the Swiss Muon Source at the Paul Scherrer Institute, Switzerland. 
\end{acknowledgments}


%

\end{document}